\documentclass[twocolumn,showpacs,amsmath,amssymb,prb,superscriptaddress]{revtex4-1}
\usepackage{graphicx}
\usepackage{color}
\begin{document}

\title{Crossover between localized states and pinned Wigner crystal in high-mobility
n-GaAs/AlGaAs heterostructures near filling factor $\nu=1$.}

\author{I.~L.~Drichko}
\author{I.~Yu.~Smirnov}
\affiliation{A.~F.~Ioffe
Physico-Technical Institute of Russian Academy of Sciences, 194021
St. Petersburg, Russia}
\author{A.~V.~Suslov}
\affiliation{National High Magnetic Field Laboratory, Tallahassee, FL 32310, USA}
\author{L.~N.~Pfeiffer}
\author{K.~W.~West}
\affiliation{Department of Electrical Engineering, Princeton University, Princeton, NJ 08544, USA}
\author{Y.~M.~Galperin}
\affiliation{Department of Physics, University of Oslo, 0316 Oslo, Norway}
\affiliation{A.~F.~Ioffe Physico-Technical Institute of Russian
Academy of Sciences, 194021 St. Petersburg, Russia}

\begin{abstract}
We have measured magnetic field dependences of the attenuation and
velocity of surface acoustic waves in a high-mobility
$n$-GaAs/AlGaAs structure with a wide quantum well. The results
allowed us to find the complex conductance, $\sigma(\omega)$, of the
heterostructure for different frequencies, temperatures and magnetic
fields near filling factors $\nu=1, 2$. Observed behavior of
$\sigma(\omega)$ versus magnetic field outside close vicinities of
integer fillings reveals an oscillation pattern indicative of the
rich fractional quantum Hall effect.

Our result is that in very close vicinities of  integer filling factors the AC response
of a high-mobility two-dimensional structures behaves as that of a
two-dimensional system of localized electrons. Namely, both real and
imaginary parts of the complex AC conductance at low temperatures
agree with the predictions for the two-site model for a
two-dimensional hopping system. Another result is the specific
temperature dependences  of  $\sigma(\omega)$, which are extremely
sensitive to the filling factor value. These dependences indicate a
sharp crossover between the localized modes and a pinned Wigner
crystal.

\end{abstract}


\maketitle

\section{Introduction}\label{Introduction}
The nature of the ground state of a two-dimensional electron system
(2DES)  in a large perpendicular magnetic field $B$ has attracted a
great interest. At small filling factors, $\nu=2\pi \hbar n/eB $,
where $n$ is the 2DES density and $e$ is the electronic charge, the
ground state in the absence of disorder is expected to be the Wigner
crystal
(WC).~\cite{Lozovik1975,Yoshioka1979,Fisher1982,Yoshioka1983,Lam1985}
Another known ground states is the fractional quantum Hall effect
(FQHE) one.~\cite{Tsui1982,Laughlin1983} Both states are induced by
electron-electron interaction. It turns out that the Laughlin FQHE
liquid states at $\nu = p/q$ (where $p$ and $q$ are integers) are
particularly robust and have ground state energies which are lower
than the WC state energy, at least for $\nu >1/5$.~\cite{Sajoto1993}
Theoretical calculations predict that, in an ideal 2DES, the WC
should be the ground state for $\nu \lesssim 1/6$. However, the WC
state may win as the filling deviates slightly from 1/5. It is
possible therefore to have a WC which is reentrant around a FQHE
liquid state, see Fig.~9 in Ref.~\onlinecite{Shayegan2005}. This
would rationalize the general current belief that the insulating
phase observed around the $\nu=1/5$ in very high quality
$n$-GaAs/AlGaAs structures is the signature of the WC state pinned
by a disorder potential. This conclusion has been confirmed using
various experimental methods. The magnetic-field-induced WC problem
in 2DESs  has been studied extensively since the late
1980's.~\cite{Shayegan1997,Pan2002}

In 2D systems along with direct current (DC) measurements of the
components of the magnetoresistance tensor,  a few research groups
study alternating current (AC) conductance $\sigma (\omega)$. The
radio-frequency electric field can be excited using the
coplanar-wave-guide technique;~\cite{Wen1969} this method was
successfully employed for studies of  the FQHE in
Ref.~\onlinecite{Engel1993} and other works.

Another probeless method of studying AC conductance uses a traveling
electric field created by a surface acoustic wave (SAW). In
connection with the integer QHE structures it was implemented in
Refs.~\onlinecite{Wixforth1986, Drichko2000} and subsequent works;
the FQHE was studied using this method in
Refs.~\onlinecite{Paalanen1992,Paalanen1992a}.

AC methods are complementary to the DC ones. In particular, specific
resonances in the AC response allow one to identify the nature of
insulating states observed at specific values of the filling factor.

Interestingly, the magnetic field dependences of $\rho_{xx}$ and
$\sigma_{xx}$ in high-mobility structures show sharp peaks (called
``wings" in Ref.~\onlinecite{Chen}) around integer values of the
filling
factor.~\cite{Pan2002,Jiang1989,Sajoto1990,Manoharan1996,Drichko2011}
These sharp wings have not ever been observed in low-mobility
systems. Numerous microwave studies  of real part,  $\sigma_1
(\omega)$, of the complex conductance, $\sigma (\omega) \equiv
\sigma_1 (\omega) -i \sigma_2 (\omega)$, close to integer $\nu$
revealed resonances at frequencies of 0.4-3~GHz, which were ascribed
to pinned modes of the Wigner
crystal.~\cite{Shayegan1997,Shayegan2005,Chen,Fertig1999, Yi2000,
Fogler2000, Chitra2001, Fogler2004,Hatke2013}

The aim of the present paper is a detailed investigation of
low-temperature complex conductivity of high-mobility 2DES in the
vicinities of integer filling factors. We use an acoustic, namely
SAW, technique to address properties of high-mobility
$n$-GaAs/AlGaAs wide quantum well at low temperatures close to
integer values of the filling factor $\nu=1$ and 2. Our main task is
study of the crossover from the electronic state at an integer $\nu$
to a pinned mode of the WC at some deviation from an integer $\nu$.
We will focus on $\nu$ close to 1, but similar behavior is also
observed close to $\nu=2$.

We will show that in high-mobility structures at low temperatures
and exactly at filling factors $\nu =1,2$  the AC response behaves
as that for localized electronic states, whereas at small deviations
from the integer values its behavior crosses over: first to that of
the Wigner crystal and then to that of the FQHE states. This
conclusion is based on experimental studies of absorption and
velocity of surface acoustic waves.

\section{Experimental method}

We use the so-called hybrid method for determining complex $\sigma
(\omega)$ from experimentally measured attenuation and velocity of a
SAW excited by interdigital transducers and propagating along a
surface of a piezoelectric  crystal (LiNbO$_3$) in a perpendicular
magnetic field. A sample containing a 2DES is mounted at the surface
of the piezoelectric crystal and is pressed to this surface by
springs. The traveling wave of electric field generated by the SAW
penetrates the 2DES causing a magnetic-field-dependent attenuation
of the SAW and change of  its velocity. The method is described in
detail in Refs.~\onlinecite{Drichko2000,Drichko2014}, in the last
paper it is sketched in Fig.~1 (left panel).

We study multilayered n-GaAlAs/GaAs/GaAlAs structures with a wide
(65 nm) GaAs quantum well (QW). The QW is $\delta$-doped from both
sides and is located at the depth d = 845 nm from the surface. The
electron density is $n=5\times 10^{10}$cm$^{-2}$  and the mobility
is $\mu_{0.3\text{K}} = 8\times 10^6$ cm$^2$/V$\cdot$s. Studies show
that at the given electron density only the lowest band of
transverse quantization should be occupied.~\cite{Manoharan1996}

\section{Results and discussion}

\begin{figure}[h!]
\centering
\includegraphics[width=0.8\columnwidth]{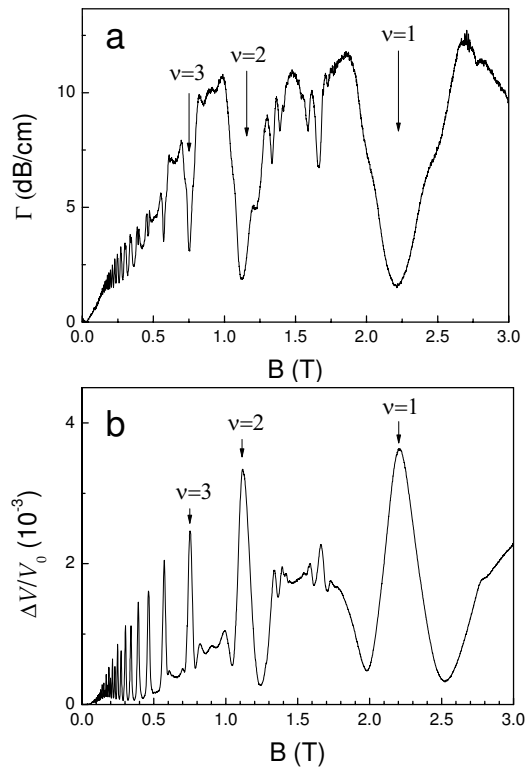}
\caption{\label{Fig2} % (Color online)
Magnetic field dependences of (a) the attenuation, $\Gamma (B)$,
and (b) SAW velocity, $\Delta V (B)/V_0$, at frequency $\omega/2\pi \equiv f=86$~MHz
at temperature $T=40$~mK}
\end{figure}

Magnetic field dependences of the attenuation, $\Gamma (B)$, and SAW
velocity, $\Delta V (B)/V_0$, were measured at temperatures of
40-380~mK and SAW frequencies of 28.5-306~MHz in magnetic fields up
to 18 T, although the analysis throughout the paper is limited by
3~T. Shown in Fig.~\ref{Fig2} are the results obtained at $T=40$~mK
and $f=86$~MHz.
These and similar dependences were used to calculate complex AC
conductance, $\sigma (\omega)$. The procedure of extracting $\sigma
(\omega)\equiv \sigma_1(\omega)-i \sigma_2(\omega)$ from the data on
$\Gamma$ and $\Delta V/V_0$ is described in detail in
Ref.~\onlinecite{Drichko2000} and references therein.

\begin{figure}[ht]
\centering
\includegraphics[width=0.8\columnwidth]{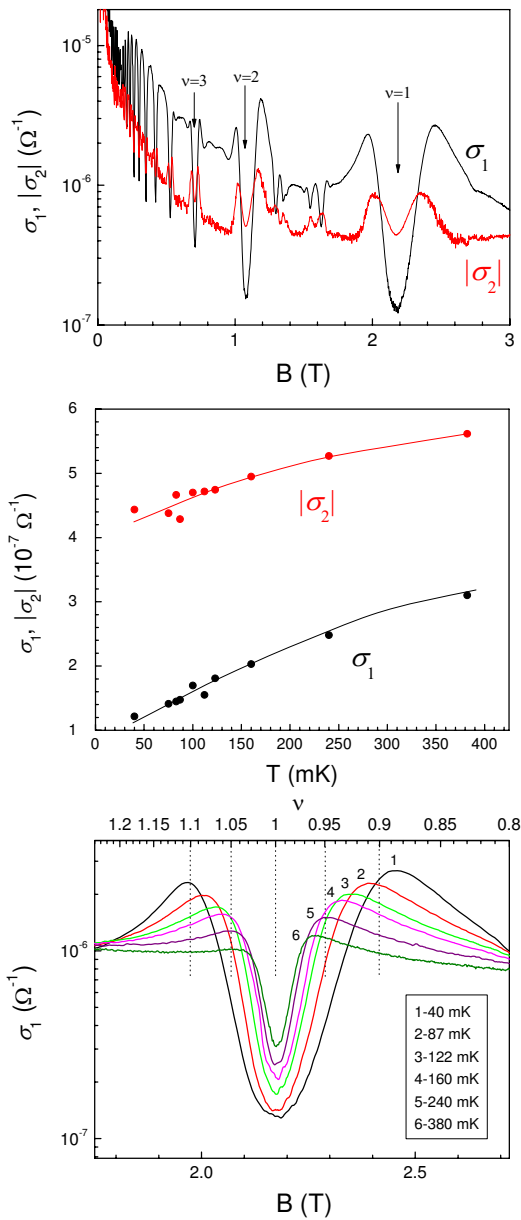}
\caption{\label{Fig3}(Color online)
Upper panel: Magnetic field dependences of $\sigma_{1}$ and
absolute value of
$\sigma_{2}$ for
$f=28.5$~MHz and $T=40$~mK. Middle panel: Temperature dependences of
$\sigma_{1}$ and $|\sigma_{2}|$ for the same frequency and $\nu =1$. The lines
are guides to the eye. Lower panel:
Magnetic field dependences of $\sigma_1$  for $\nu$ close to 1 at various
temperatures.}
\end{figure}

Magnetic field dependences of $\sigma_{1}$ and $|\sigma_{2}|$ for
$f=28.5$~MHz and $T=40$~mK are shown in the upper panel of
Fig.~\ref{Fig3}. The sign of $\sigma_2(\omega)$ will be discussed
later, see the discussion related to Fig.~\ref{Fig5}.
Shown in the middle panel are the temperature
dependences of $\sigma_{1}$ and $|\sigma_{2}|$ for the same frequency
and $\nu =1$. In the lower panel, the magnetic field dependences of
$\sigma_{1}$ and $|\sigma_{2}|$ for $\nu$ close to 1 at different
temperatures are shown.

One can see that integer values of $\nu$ correspond to the minima in
$\sigma_1(\omega)$. Close to the minima $|\sigma_2 (\omega)| >
\sigma_1 (\omega)$. Between these minima rich oscillation patterns
typical for the FQHE were observed. Interestingly, the minima and
the FQHE regions are separated by sharp maxima (``wings") - each
minimum is surrounded by two wings. Outside close vicinities of
integer $\nu$, $|\sigma_2 (\omega) |< \sigma_1 (\omega)$. The
behavior described above is observed only in high-mobility samples.
In the low-mobility samples, pronounced maxima of $\sigma_1(\omega)$
at half-integer $\nu$ are observed rather than the FQHE related
oscillations. Consequently, the wings around integer $\nu$ values
are absent.

Let us now consider the behavior of $\sigma(\omega)$ at  `exactly'
 integer $\nu$. We focus on temperature dependences of $\sigma_{1}$
and $|\sigma_{2}|$ at $\nu =1$ shown in the middle panel of
Fig.~\ref{Fig3}. One observes that at $T<400$~mK $\sigma_1$
\textit{increases} with temperature, and in this temperature domain
$|\sigma_2|
>\sigma_1$ in agreement with Ref.~\onlinecite{EfrosTSM}.
The temperature dependence of $\sigma_1$ at $\nu=2$ is similar.

The frequency dependence of $\sigma_1$ in the above temperature
domain is weak: As the frequency changes by factor 11, $\sigma_1$
changes only by 20\%. The values of $\sigma_1$ at different minima
corresponding to integer $\nu$ decrease with magnetic field $\propto
B^{-1.8}$ as it follows from the analysis for $\nu =2,4,6,8$, and
10. This dependence is compatible with theoretical prediction
$\sigma_1 (\omega) \propto B^{-2}$  based on the two-site model for
absorption by localized states.~\cite{Galperin1986}

Now let us recall that in low (medium) mobility  systems showing
only an integer quantum Hall effect behavior of $\sigma_1$ is well
described by the one-electron picture involving electrons trapped by
a random potential. According to this picture, at integer $\nu$ the
Fermi level is located in the middle of the distance between the
Landau levels, the electron states are localized by disorder, and
low-temperature DC conductance, $\sigma_{\text{DC}}$, is
exponentially small. The AC conductance is determined by electron
hops between nearest potential minima resulting in $\sigma_1(\omega)
\gg \sigma_{\text{DC}}$. In this case, the AC response can be
explained by the two-site model, for a review see
Refs.~\onlinecite{Efros1985,Galperin1991} and references therein.
According to this model, a pair of the electron energy minima is
described as a two-level tunneling system (TLS) with diagonal
splitting $\Delta$ and tunneling splitting $\Lambda (r)$, the
interlevel spacing being $E=\sqrt {\Delta^2 + \Lambda^2}$. At
sufficiently low frequencies the AC response is due to relaxation of
the nonequilibrium populations of the minima. The corresponding
relaxation rate can be expressed as, cf.
Ref.~\onlinecite{Galperin1991},
\begin{equation}\label{r-time}
\frac{1}{\tau (E,r)}=\frac{1}{\tau_0 (T)}F\left(\frac{E}{kT}\right)\left(\frac{\Lambda(r)}{E}\right)^2\, .
\end{equation}
Here $k$ is the Boltzmann constant. Equation \eqref{r-time} assumes
that the levels' populations relax due to interaction between
localized electrons and phonons.  The interaction matrix element
contains, therefore, the electron-phonon coupling constant as well
as the tunneling coupling between the sites, $\Lambda (r)$, which
exponentially decays with the distance $r$ between the minima.
The corresponding relaxation rate, $\tau^{-1} (E,r)$, is, therefore, proportional to $\Lambda^2 (r)$ that is taken into account by the last factor in Eq.~\eqref{r-time}. In addition to the coupling constant squared, the rate is proportional to the  phonon density of states at frequency  $E/\hbar$. However, only the configurations with $E\lesssim kT$ are important since the configurations with $E \gg kT$ are frozen in their ground states. Therefore, we split the
rate into the factor $\tau_0^{-1} (T)$ (corresponding to the systems with $E=kT$) and dimensionless function, $F$, which depends on the details of the electron-phonon interaction.~\cite{Galperin1991} It is normalized in order to have $F(1)=1$. Since $\Lambda(r) \le E$, the time $\tau_0$ has a meaning of the
\textit{minimal} relaxation time for a TLS with the level splitting
$E=kT$.

The theory predicts that (with logarithmic accuracy)~\cite{Galperin1991}
\begin{equation}\label{rel2}
\sigma_1(\omega) \propto \min\{\omega, \tau_0^{-1}(T)\}\,, \quad |\sigma_2(\omega)| \gtrsim \sigma_1(\omega) \, .
\end{equation}
The first of the above expressions allows a simple qualitative
interpretation.  Let us consider TLSs with $E\approx kT$, which
plays the main role.   The contribution of a TLS to the dissipation
depends on the product $\omega \tau$. Very ``fast" systems with
$\tau \ll  \omega^{-1}$ do not essentially contribute because their
populations almost adiabatically follow the AC electric field. On
the other hand, very ``slow" systems having $\tau \gg \omega^{-1}$
also do not contribute since their populations have not enough time
to follow the AC field. Therefore, the optimal ones are those having
$\tau \sim \omega^{-1}$.

On the other hand, since $\Lambda (r)$ exponentially decreases with
$r$  there exists an exponentially broad  set of systems having
relaxation times longer than $\tau_0 (T)$. Therefore, if $\omega
\tau_0 \ll 1$ the optimal pairs  with $\omega \tau \sim 1$ can
always be found, and  it is those pairs that provide the main
contribution to the absorption. On the contrary, at $\omega \tau_0
\gg 1$ the optimal pairs are absent, and the absorption is dominated
by the pairs with $\tau \sim \tau_0$.

Looking at our data we
 conclude that at $\nu =1$ the behavior of $\sigma_{1}$ and
$\sigma_{2}$ in our sample is compatible with the picture of
relaxation absorption of SAW by localized electrons under condition
$\omega \gg \tau_0^{-1}$, see Eq.~(\ref{rel2}). Indeed, estimates
based on Eq.~(\ref{r-time}) show that the main contribution to the
relaxation rate $\tau_0^{-1}$ is due to piezoelectric interaction
between localized electrons and phonons. In this case, see, e.~g.,
Ref.~\onlinecite{Galperin1991}, $\tau_0^{-1}(T)$ is roughly
proportional to $T$, and at 40$~\mathrm{mK}$ $\tau_0=1.4\cdot
10^{-8}$~s.

Now let us discuss the magnetic field dependences of $\sigma_{1}$ in
the vicinity of $\nu =1$, the behavior of $\sigma_{1}$ around $\nu =
2$ being similar. As is seen in the lower panel of Fig.~\ref{Fig3},
the minimum of $\sigma_1$ corresponding to $\nu=1$ is surrounded by
maxima heights and locations of these maxima depend on temperature.
\begin{figure}[t]
\centering
\includegraphics[width=.9\columnwidth]{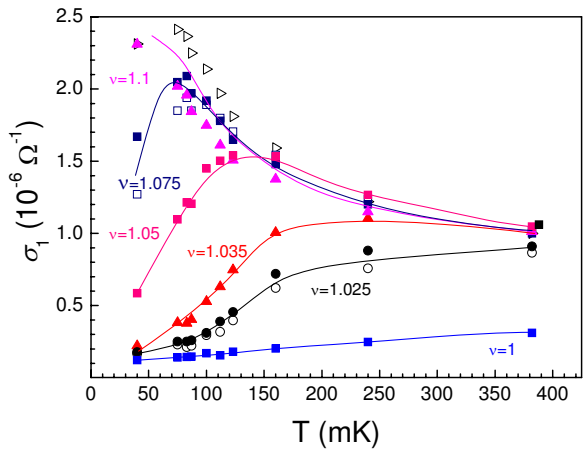}
\caption{\label{Fig4}
(Color online)
Temperature dependences of $\sigma_1 (28.5~\mathrm{MHz})$ for
different values of $\nu$ (shown near the curves). Mainly, the
results for $\nu \ge 1$ are demonstrated (the filled symbols); the
picture is almost symmetric for $\nu<1$ as partially shown by
open symbols $\rhd$ ($\nu$=0.9), $\square$ ($\nu$=0.925) and $\bigcirc$ ($\nu$=0.975).
The lines are guides to the eye.}
\end{figure}

Temperature dependences of $\sigma_1$ at frequency 28.5 MHz for
different values of the filling factor in the range 0.9 $\leq \nu
\leq$ 1.1 are shown in Fig.~\ref{Fig4}. The dependences obtained for
other investigated SAW frequencies are similar. Let us first
consider these temperature dependences at the range ends, i.e., at
$\nu$= 1.1 and 0.9. We conclude that the electronic state at
$\nu$=1.1(0.9) is indeed the WC. This conclusion came from (i)
dramatic increase of the conductance as against $\nu$= 1; (ii)
different temperature dependences of  $\sigma_1$ at   $\nu$= 1 and
$\nu$ = 1.1 (0.9), (these dependences at = 1.1 and 0.9 are similar
to those at  $\nu$ = 0.19 and 0.21, whereas the formation of Wigner
solids in the latter cases were proved by various authors who used a
number of experimental techniques;~\cite{Sajoto1993,Pan2002}) and
(iii) the frequency dependence of $\sigma_2 (\omega)$ shown in
Fig.~\ref{Fig5} for different values of the filling factor, which
demonstrates zero crossing of the  $\sigma_2 (\omega)$ at $\nu$= 1.1
at some frequency that is typical for the Wigner crystal. According
to Ref.~\onlinecite{Hatke2013}, the frequency corresponding to the
zero crossing is equal to the pinning frequency of the Wigner solid.
This zero crossing in the frequency dependence of the imaginary part
$\sigma_2 (\omega)$ should be accompanied by a maximum in the
frequency dependence of the real part of $\sigma_1 (\omega)$. We did
not observe such maxima due to precision limitations of our
equipment.
\begin{figure}[h!]
\centering
\includegraphics[width=.9\columnwidth]{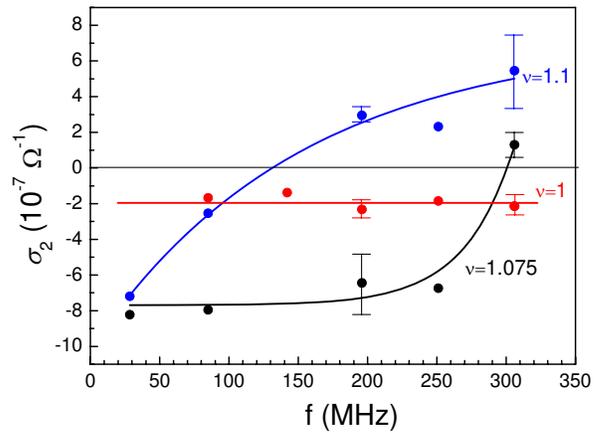} \\
\
\caption{\label{Fig5} (Color online)
Frequency dependences of $\sigma_2$ for different $\nu$. $T=40$~mK.
}
\end{figure}

Studies based on microwave spectroscopy, see, e.~g.,
Ref.~\onlinecite{Chen} and references therein, also lead to the
conclusion that for  $\nu$ = 1.1 and 0.9 the electron state in a
similar sample can be understood as a Wigner solid. This conclusion
was based on observation of resonances in the frequency dependence
of $\sigma_1$.

Let us now analyze the temperature dependences of $\sigma_1$ inside
the filling factor range 0.9 $\leq \nu \leq$ 1.1. For all filling
factor values at low temperatures conductivity $\sigma_1$ initially
rises as the temperature increases. However, as the conductivity
reaches the value corresponding  $\nu$= 1.1 (0.9) its temperature
dependence changes, and the conductivity begins to decrease as the
temperature rises.

Notice that the characteristic temperature of this crossover in the
temperature dependence of conductivity decreases with increasing
deviation of the filing factor from the unity,  $|\nu -1|$.

We assume that the observed initial increase of  $\sigma_1$ with
rising temperature at all values   $\nu <$ 1.1 is associated with
the hopping nature of this conductivity at low temperatures.
Assuming that the conductance of the localized phase can be
represented by the single-electron expression (\ref{rel2}) at
$\omega \tau_0 \gg 1$ we expect that $\sigma_1 (\omega) \propto
\tau_0^{-1}(T)$ is an increasing function of temperature, with the
slope proportional to the squared single-electron density of states,
$g^2$.~\cite{Galperin1991} Therefore, as the value $ |\nu -1|$
increases in the range $|\nu -1| \leq 0.1$ both the single-electron
density and the slope rise.

As the conductivity is increasing, conditions favorable for formation of
the Wigner crystal are emerging and at some temperature the crystal
gets formed. The characteristic temperature of this crossover is
decreasing with increasing deviation of the filing factor from an
integer value, in this case  $| \nu - 1 |$.

\begin{figure}[h!]
\centering
\includegraphics[width=.9\columnwidth]{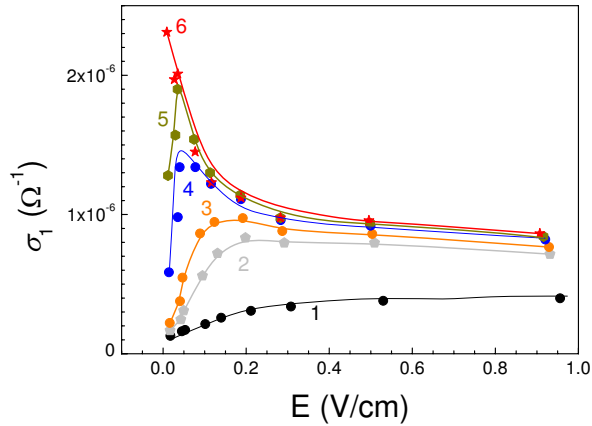} \\
\
\caption{\label{Fig6} (Color online)
Dependence of the real part of the conductivity $\sigma_1$ on the
SAW electric field near $\nu$=1: 1 - $\nu$=1.0, 2 - $\nu$=1.025, 3 - $\nu$=1.035,
4 - $\nu$=1.05, 5 - $\nu$=1.075, 6 - $\nu$=1.1; $T$=40 mK, $f$=28.5 MHz.
The lines are guides to the eye.
}
\end{figure}

We have also conducted measurements of the SAW intensity impact on
the ac conductance near $\nu$=1.  In Fig.~\ref{Fig6} we present the
dependence of the real part of AC conductivity, $\sigma_1$, on the
electric field, accompanying the SAW of frequency 28.5 MHz at
temperature 40 mK. The dependences obtained at other frequencies are
similar.  The electric field was calculated from values of SAW
intensity using Eq.~(6) of Ref.~\onlinecite{Drichko2011}. The
dependences $\sigma_1(E)$ are qualitatively similar to the
dependences $\sigma_1 (T)$ shown in Fig.~\ref{Fig4}. Therefore,
 increase of SAW intensity acts as increase of temperature -- the SAW heats the electron system.
This mechanism differs from the predicted one based on nonlinear sliding of the WC.\cite{Normand1992,*Zhu1994}
Assuming that the electron system can be characterized by an effective temperature, $T_e$,
and performing analysis similar to the one described in Ref.~\onlinecite{Drichko2011} for $\nu = 1.1$ we
conclude that the heat released is proportional to $(T_e^3 -T^3)$.
It is worth noting that when the temperature (or the SAW intensity)
increases the behavior of the complex conductance can be interpreted
as manifestation of WC melting.

For $\nu = 1$, the dependence $\sigma_1 (E)$ is increasing, but rather weak.
Therefore, it is had to identify the underlying mechanism of nonlinearity. We attribute
the observed increase of AC conductivity with $E$ to a combination of electron heating
and field-induced ionization of the electron states localized in shallow wells of random
potential.

In this way we arrive at the following scenario. In high-mobility
2DES the electronic states at small integer filling factors 1 and 2
are localized. However, very small deviation from integer $\nu$
leads to delocalization facilitating formation of a collective mode
- a pinned WC - due to pronounced electron-electron interaction.
Further deviation from integer $\nu$ results in formation of FQHE
states.

\section{Conclusion}
%\paragraph{Conclusion --}

1. In this paper we have measured magnetic field dependences of the
attenuation and velocity of a SAW in high-mobility n-GaAs/AlGaAs
structure. The results allowed us finding complex conductance,
$\sigma (\omega)\equiv \sigma_1(\omega)-i \sigma_2(\omega)$, for
different frequencies, temperatures and magnetic fields.

2. We found that at a small exact integer filling factor ($\nu$=1
and 2) 2D electrons were localized. The observed AC conductivity is
of the hopping nature and it agrees with the two-site model provided
that $\omega \tau_0 > 1$.

3. We also found that at filling factor  $\nu$=1.1 a pinned Wigner
solid was formed.

4. At small deviations of the filling factor from exact integers
sharp crossovers between the localized states and Wigner solids were
observed in the temperature dependences of $\sigma_1$.

\acknowledgments
%\section*{Acknowledgments}

%\paragraph{Acknowledgment --}  (if needed)
I.L.D. is grateful for support from Russian Foundation for Basic
Research via grant 14-02-00232. The authors would like to thank E.
Palm, T. Murphy, J.-H. Park, and G. Jones for technical assistance.
NHMFL is supported by National Science Foundation Cooperative
Agreement No. DMR-1157490 and the State of Florida. The work at
Princeton was partially funded by the Gordon and Betty Moore
Foundation through Grant GBMF2719, and by the National Science
Foundation MRSEC-DMR-0819860 at the Princeton Center for Complex
Materials.

%\bibiographystyle{jetpl}
%bibliography{wigner}
%\end{document}
%\input{Nu_one.bbl}

\end{document}